Special features of the KdV-Sawada-Kotera equation

Yair Zarmi
Jacob Blaustein Institutes for Desert Research
Ben-Gurion University of the Negev
Midreshet Ben-Gurion, 84990, Israel

ABSTRACT

The KdV-Sawada-Kotera equation has single-, two- and three-soliton solutions. However, it is not known yet whether it has *N*-soliton solutions for any *N*. Viewing it as a perturbed KdV equation, the asymptotic expansion of the solution is developed through third order within the framework of a Normal Form analysis. It is shown that the equation is asymptotically integrable through the order considered. Focusing on the soliton sector, it is shown that the higher-order corrections in the Normal Form expansion represent purely inelastic KdV-soliton-collision processes, and vanish identically in the single-soliton limit. These characteristics are satisfied by the exact two-soliton solution of the KdV-Sawada-Kotera equation: The deviation of this solution from its KdV-type two-soliton approximation describes a purely inelastic scattering process: The incoming state is the faster KdV soliton. It propagates until it hits a localized perturbation, which causes its transformation into the outgoing state, the slower soliton. In addition, the effect of the perturbation on the exact two-soliton solution vanishes identically in the single-soliton limit (equal wave numbers for the two solitons).

## 1. The KdV-Sawada-Kotera equation

The KdV equation with a Sawada-Kotera [1] perturbation,

$$w_t = 6ww_1 + w_3 + \varepsilon\alpha_4\left(45w^2 w_1 + 15ww_3 + 15w_1 w_2 + w_5\right) \quad \left(w_k \equiv \partial_x^k w\right) , \tag{1}$$

has been investigated extensively in the literature [2-8]. It is integrable in the Painlevé sense [2, 6], and has 1- 2- and 3-soliton solutions [2-8]. They all have the Hirota structure [9]

$$w = 2\partial_x^2 \ln f(t,x) . \tag{2}$$

In the case of the single-soliton solution, $f(x)$ is given by

$$f(t,x) = 1 + g , \tag{3}$$

with

$$g = \exp\left[2k(x + v(k)t)\right] , \quad v(k) = 4k^2\left(1 + \varepsilon\alpha_4 4k^2\right) . \tag{4}$$

The single-soliton solution is:

$$u^{Single}(t,x;k) = \frac{2k^2}{\cosh\left[k(x + v(k)t)\right]^2} . \tag{5}$$

In the two-soliton case, one has

$$f(t,x) = 1 + g_1 + g_2 + A g_1 g_2 . \tag{6}$$

In Eq. (6), $g_i$ are defined by Eq. (4), with $k \to k_i$, $i = 1, 2$, and

$$A = \frac{(k_1 - k_2)^2}{(k_1 + k_2)^2} \frac{\left(1 + \frac{20}{3}\varepsilon\alpha_4\left(k_1^2 - k_1 k_2 + k_2^2\right)\right)}{\left(1 + \frac{20}{3}\varepsilon\alpha_4\left(k_1^2 + k_1 k_2 + k_2^2\right)\right)} . \tag{7}$$

Whether Eq. (1) has an infinite family of multiple-soliton solutions, is not clear yet. However, treating Eq. (1) as a perturbed KdV equation, it is shown to be integrable asymptotically for *any* zero-order approximation, at least through $O(\varepsilon^3)$. This is achieved through the construction of its solution within the framework of a Normal Form asymptotic expansion [10-14] in Section 2.

The exact 1- and 2-soliton solution solutions of Eq. (1) are then used as "templates", against which the Normal Form expansion of the solution in the multiple-soliton case is compared. The following properties of the exact solutions are made use of. First, the single-soliton solution, given in Eq. (5), is identical to that of a single-KdV-soliton solution of the Normal Form. The only effect of the perturbation on this solution is updating of the soliton velocity, $v(k)$, as shown in Eq. (4).

The properties of the exact two-soliton solution, given by Eqs. (2), (4), (6) and (7) are: First, in the single-soliton limit ($k_2 = k_1$), this solution degenerates into the single-soliton solution of Eq. (5), so that the effect if the perturbation disappears. Second, the asymptotic structure of the two-soliton solution describes the elastic collision of two KdV-type solitons, the only effect of the perturbation being a modification of the phase shifts beyond their KdV value (denoted by $\delta_{i,0}$). For $k_2 > k_1$, Eqs. (2), (4), (6) and (7) yield

$$w^{Two-solitons} \rightarrow \begin{cases} u^{Single}(t,x;k_1) + u^{Single}(t,x+\delta_2;k_2) &, \quad t \rightarrow -\infty \\ u^{Single}(t,x;k_2) + u^{Single}(t,x+\delta_1;k_1) &, \quad t \rightarrow +\infty \end{cases}. \tag{8}$$

In Eq. (8),

$$\delta_i = \delta_{i,0} + \frac{1}{k_i} \ln\left[\sqrt{\frac{1 + \frac{20}{3}\varepsilon a_4\left(k_1^2 - k_1 k_2 + k_2^2\right)}{1 + \frac{20}{3}\varepsilon a_4\left(k_1^2 + k_1 k_2 + k_2^2\right)}}\right] \quad \left(\delta_{i,0} = \frac{1}{k_i}\ln\left[\frac{k_2 - k_1}{k_1 + k_2}\right]\right). \tag{9}$$

The significance of this result is that, if one subtracts from the exact two-soliton solution its zero-order KdV-approximation, $u(t,x)$ (obtained by setting $\varepsilon = 0$ in $A$ of Eq. (7)), then, in both incoming and outgoing states, the solitons without phase shifts disappear completely. Thus, this difference represents a purely inelastic scattering process of KdV solitons. The incoming state in the differ-ence is the faster soliton. It propagates until it hits a localized "scattering potential", which "absorbs" it, and "emits" the slower soliton as the outgoing state. This inelastic scattering process, represented by the difference $(w(t,x) - u(t,x))/\varepsilon$, is shown in Fig. 2.

The higher-order terms in the Normal Form expansion are analyzed in Section 3. The aspects that characterize the exact single- and two-soliton solutions of Eq. (1) are obeyed by the Normal-Form asymptotic expansion of the solution when $u$, the zero-order approximation, is *any* multiple-soliton solution. First, a solution exists, in which the higher-order correction terms vanish identically in the single-soliton limit; namely, the higher-order terms in Eq. (9) vanish if $u^{Single}$ is substituted for $u$. Whether the higher-order terms do or do not vanish in the single-soliton limit for other solutions of Eq. (1) depends on the initial data and boundary conditions imposed on the solution.

The second aspect is the nature of soliton collision processes described by the higher-order corrections in the multiple-soliton case. If in Eq. (1) the coefficients of the various monomials are replaced by arbitrary values, then, in a Normal Form analysis, the higher-order corrections contain contributions that correspond to elastic soliton collisions (terms that asymptote into well separated single-soliton contributions, which are unaffected by the existence of other solitons) and inelastic collisions (terms that asymptote into well separated single-soliton contributions, which *are* affected by the existence of other solitons) [15]. In the case of Eq. (1), the higher-order corrections to its solutions represent purely inelastic processes.

**2. Asymptotic integrability**
Eq. (1) is a particular case of the KdV equation, to which a first-order perturbation is added:

$$w_t = 6 w w_1 + w_3 + \varepsilon \left( 30 \alpha_1 w^2 w_1 + 10 \alpha_2 w w_3 + 20 \alpha_3 w_1 w_2 + \alpha_4 w_5 \right) \ . \tag{10}$$

In the Normal Form analysis [10-14], one expands the solution of Eq. (10) in a power series in $\varepsilon$:

$$w = u + \varepsilon u^{(1)} + \varepsilon^2 u^{(2)} + \varepsilon^3 u^{(3)} + O(\varepsilon^4) \ . \tag{11}$$

The zero-order approximation, $u(t,x)$, is expected to be governed by an integrable Normal Form [10-14]. As the perturbation contains only an $O(\varepsilon)$ contribution, the Normal Form is a sum of two terms:

$$u_t = 6uu_1 + u_3 + \varepsilon\alpha_4\left(30u^2 u_1 + 10uu_3 + 20u_1 u_2 + u_5\right) \ . \tag{12}$$

Eq. (12) has the same infinite family of $N$-soliton solutions, $N = 1, 2, 3,\ldots$, as the unperturbed KdV equation, with soliton velocities updated by the effect of the perturbation as in Eq. (4) [10-14].

In addition, it has been hoped that $u^{(k)}$, $k \geq 1$, could be solved for in terms of differential polynomials in $u$ (polynomials in $u$ and its spatial derivatives). However, whereas $u^{(1)}$ can be solved for in closed form [10], in general, $u^{(k)}$, $k \geq 2$, cannot. It is impossible to express $u^{(k)}$ as differential polynomials [11-14] unless the coefficients in Eq. (10) obey algebraic constraints. The conditions for $O(\varepsilon^2)$ and $O(\varepsilon^3)$ asymptotic integrability are [11-15]:

$$\mu_2 = 3\alpha_1\alpha_2 + 4\alpha_2^2 - 18\alpha_1\alpha_3 + 60\alpha_2\alpha_3 - 24\alpha_3^2 + 18\alpha_1\alpha_4 - 67\alpha_2\alpha_4 + 24\alpha_4^2 = 0 \ . \tag{13}$$

and

$$\begin{aligned}\mu_{31} =\ & \tfrac{100}{3}\alpha_1^2\alpha_2 + \tfrac{325}{9}\alpha_1\alpha_2^2 + \tfrac{500}{27}\alpha_2^3 - 200\alpha_1^2\alpha_3 + \tfrac{4300}{9}\alpha_1\alpha_2\alpha_3 + \tfrac{5500}{27}\alpha_2^2\alpha_3 - \tfrac{500}{3}\alpha_1\alpha_3^2 \\ & - \tfrac{2000}{9}\alpha_2\alpha_3^2 + \tfrac{2000}{27}\alpha_3^3 + 150\alpha_1^2\alpha_4 - \tfrac{2930}{9}\alpha_1\alpha_2\alpha_4 - \tfrac{12895}{27}\alpha_2^2\alpha_4 + \tfrac{1030}{9}\alpha_1\alpha_3\alpha_4 \\ & - \tfrac{2600}{9}\alpha_2\alpha_3\alpha_4 + \tfrac{1040}{9}\alpha_3^2\alpha_4 - \tfrac{260}{3}\alpha_1\alpha_4^2 + \tfrac{20210}{27}\alpha_2\alpha_4^2 - \tfrac{1840}{9}\alpha_4^3 = 0\end{aligned} \tag{14}$$

$$\begin{aligned}\mu_{32} =\ & -\tfrac{350}{3}\alpha_1^2\alpha_2 - \tfrac{1100}{9}\alpha_1\alpha_2^2 - \tfrac{2050}{27}\alpha_2^3 + 700\alpha_1^2\alpha_3 - \tfrac{14300}{9}\alpha_1\alpha_2\alpha_3 - \tfrac{24500}{27}\alpha_2^2\alpha_3 + 600\alpha_1\alpha_3^2 \\ & + \tfrac{5600}{9}\alpha_2\alpha_3^2 - \tfrac{4000}{27}\alpha_3^3 - 500\alpha_1^2\alpha_4 + \tfrac{31015}{36}\alpha_1\alpha_2\alpha_4 + \tfrac{115555}{54}\alpha_2^2\alpha_4 - \tfrac{9295}{18}\alpha_1\alpha_3\alpha_4 \\ & + \tfrac{16325}{9}\alpha_2\alpha_3\alpha_4 - \tfrac{6530}{9}\alpha_3^2\alpha_4 + \tfrac{3265}{6}\alpha_1\alpha_4^2 - \tfrac{386755}{108}\alpha_2\alpha_4^2 + \tfrac{9005}{9}\alpha_4^3 = 0\end{aligned} \tag{15}$$

$$\begin{aligned}\mu_{33} =\ & -\tfrac{40}{3}\alpha_1^2\alpha_2 - \tfrac{100}{9}\alpha_1\alpha_2^2 - \tfrac{290}{27}\alpha_2^3 + 80\alpha_1^2\alpha_3 - \tfrac{1120}{9}\alpha_1\alpha_2\alpha_3 - \tfrac{5500}{27}\alpha_2^2\alpha_3 + 80\alpha_1\alpha_3^2 \\ & - \tfrac{320}{9}\alpha_2\alpha_3^2 + \tfrac{1600}{27}\alpha_3^3 - 40\alpha_1^2\alpha_4 - \tfrac{1621}{36}\alpha_1\alpha_2\alpha_4 + \tfrac{22463}{54}\alpha_2^2\alpha_4 - \tfrac{2627}{18}\alpha_1\alpha_3\alpha_4 \\ & + \tfrac{5545}{9}\alpha_2\alpha_3\alpha_4 - \tfrac{2218}{9}\alpha_3^2\alpha_4 + \tfrac{1109}{6}\alpha_1\alpha_4^2 - \tfrac{81503}{108}\alpha_2\alpha_4^2 + \tfrac{1753}{9}\alpha_4^3 = 0\end{aligned} \tag{16}$$

Eq. (1) is a special case of Eq. (10) with

$$\alpha_1 = \frac{3}{2}\alpha_4 \quad , \quad \alpha_2 = \frac{3}{2}\alpha_4 \quad , \quad \alpha_3 = \frac{3}{4}\alpha_4 \quad . \tag{17}$$

Substitution of Eq. (17) in Eqs. (13)-(16) reveals that $\mu_2$, $\mu_{31}$, $\mu_{32}$ and $\mu_{33}$ all vanish. Thus, Eq. (1) *is* asymptotically integrable at least through $O(\varepsilon^3)$. Hence, not only $u^{(1)}$, but also $u^{(2)}$, and $u^{(3)}$ can be solved for in terms of differential polynomials in $u$.

### 3. Higher-order corrections are purely inelastic and vanish in single-soliton limit
### 3.1 Normal Form derivation of $u^{(1)}$

To emphasize the special characteristics of Eq. (1), let us return to the Normal Form analysis [10-14] of the case with a general perturbation, Eq. (10). Substituting Eqs. (11) and (12) in Eq. (10), the $O(\varepsilon)$-equation for $u^{(1)}$ is found to be:

$$\partial_t u^{(1)} = 6\partial_x \left(u u^{(1)}\right) + \partial_x^3 u^{(1)} + \left(30(\alpha_1 - \alpha_4)u^2 u_1 + 10(\alpha_2 - \alpha_4)u u_3 + 20(\alpha_3 - \alpha_4)u_1 u_2\right) . \tag{18}$$

That $u^{(1)}$ can be solved for in closed form has been known for years [10]. It is instructive to write this solution as follows [15]:

$$u^{(1)} = u_{el}^{(1)} + u_{in}^{(1)} + S \quad , \tag{19}$$

where

$$u_{el}^{(1)} = \left(-\tfrac{5}{2}\alpha_1 + \tfrac{10}{3}\alpha_2 + \tfrac{5}{3}\alpha_3 - \tfrac{5}{2}\alpha_4\right)u_2 + \left(-5\alpha_1 + 5\alpha_2\right)u^2 \quad , \tag{20}$$

and

$$u_{in}^{(1)} = -\tfrac{10}{3}(\alpha_2 - \alpha_4)\partial_x(u_1 + q u) \quad , \quad q = \partial_x^{-1} u = \int^x u(t,x)dx \quad . \tag{21}$$

In Eq. (19), $S$ is a solution of the homogeneous version of Eq. (18) – a linear combination of symmetries of the KdV equation [10-14,16-23]. It is required solely for the purpose of satisfying the initial data and boundary conditions imposed on $u^{(1)}$. $u_{el}^{(1)}$ represents elastic soliton scattering. When $u$ is a multiple-soliton solution of the Normal Form, $u_{el}^{(1)}$ asymptotes into a sum of well-separated single-soliton solutions of the Normal Form. Apart from the standard KdV phase shifts, each of the solitons is unaffected by the existence of the other solitons. In contradistinction, $u_{in}^{(1)}$ represents inelastic soliton scattering. It asymptotes into a sum of well-separated single-soliton solutions of the Normal Form, each of which *is* affected by the existence of the other solitons.

Let us now return to Eq. (1). With the values of $\alpha_i$, $1 \leq i \leq 3$ given in Eq. (17)) the elastic component, $u_{el}^{(1)}$, vanishes identically. Namely, for any multiple-soliton zero-order approximation, $u^{(1)}$ represents a purely inelastic process. In addition, Eq. (18) becomes

$$u_t^{(1)} = 6\left(u\, u^{(1)}\right)_x + u_{xxx}^{(1)} + 5\alpha_4 \partial_x R[u] \quad , \quad \left(R = u^3 - u_1^2 + u\, u_2\right) \ . \tag{22}$$

The driving term, $R$ is a *local special polynomial* [15]: It vanishes identically when $u$ is the single-soliton solution of Eq. (5), and contains only terms that are localized around the soliton trajectories in the multiple-soliton case. It represents a genuine soliton interaction term because it is localized around the soliton collision region in the multiple-soliton case. Eq. (19) for $u^{(1)}$ now becomes:

$$u^{(1)} = u_{in}^{(1)} + S \quad , \quad u_{in}^{(1)} = -\frac{5}{3}\alpha_4 \partial_x \left(u_1 + q\, u\right) \ . \tag{23}$$

As the driving term in Eq. (22) vanishes identically in the single-soliton case, it is desirable to ensure that the particular solution, $u_{in}^{(1)}$, also vanishes in that case. This can be achieved by observing that the usual definition of the non-local entity, $q(t,x)$, in Eq. (21) does not specify both limits of integration. The reasoning has been that only the upper limit is important. The following choice of integration limits,

$$q(t,x) = \frac{1}{2}\left\{\int_{-\infty}^{x} u(t,x)dx - \int_{x}^{\infty} u(t,x)dx\right\} = \int_{-\infty}^{x} u(t,x)dx - \frac{1}{2}\int_{-\infty}^{\infty} u(t,x)dx \ , \tag{24}$$

ensures that $u_{in}^{(1)}$ vanishes in the single-soliton case [15]. $u_{in}^{(1)}$ is then a *non-local special polynomial* because it contains the non-local entity, $q(t,x)$. (Other choices of the limits of integration merely modify $S$, the linear combination of symmetries of the KdV equation.)

With the definition of Eq. (24), $u_{in}^{(1)}$ has a simple structure when the zero-order term, $u$ is an $N$-soliton solution of the Normal Form, Eq. (12). The latter is just an $N$-soliton solution of the KdV equation, with the velocity of each soliton given by [10-14]:

$$v(k_i) = 4 k_i^2 \left(1 + \varepsilon \alpha_4 \, 4 k_i^2 \right) . \tag{25}$$

Denoting the standard KdV phase shifts by $\delta_{i,0}$, the asymptotic form of $u_{in}^{(1)}$ is [15]:

$$u_{in}^{(1)}(t,x) \xrightarrow[|t|\to\infty]{} -\frac{5}{3}\alpha_4 \, \partial_x \left\{ \sum_{i=1}^{N} Q_i^{(1)} u^{Single}\left(t, x + \delta_{i,0}; k_i\right) \right\} , \tag{26}$$

where

$$\left( Q_i^{(1)} = \left\{ \sum_{k_j < k_i}(-2 k_j) + \sum_{k_j > k_i}(2 k_j) \right\} \mathrm{sgn}(t) \right) . \tag{27}$$

Thus, apart a solution of the homogeneous version of Eq. (18), $u_{in}^{(1)}$ asymptotes into a sum of well separated single KdV-type soliton solutions of the Normal Form, Eq. (12). It represents a *pure inelastic* scattering process: The amplitude of each soliton is multiplied by wave numbers of *other* solitons. In the two-soliton case, Eq. (26) describes collision between a soliton and an anti-soliton (negative amplitude soliton), which exchange signs upon collision.

**3.2 First-order correction in exact two-soliton solution of Eq. (1)**
To obtain $u^{(1)}$, the full first-order correction to the solution of Eq. (1) in the Normal Form expansion, one must specify $S$, the linear combination of symmetries of the KdV equation, which has to be added in Eq. (23) to the particular solution, $u_{in}^{(1)}$. The structure of $S$ depends on the specific so-lution of Eq. (1) that is selected. As an example, let us return to the exact two-soliton solution of Eq. (1), given by Eqs. (2), (4), (6), (7) and (25). The first step is the expansion of this solution in powers of $\varepsilon$, so as to obtain Eq. (11). The expansion ought to account only for the appearance of $\varepsilon$ in the coefficient $A$ of Eq. (7). (The $\varepsilon$-dependence in the velocities (Eq. (22)) must be left intact, as the velocities *are* updated by the effect of the perturbation in the Normal Form.)

The zero-order term, $u$, is the two-soliton solution of the Normal Form, Eq. (12). The first-order correction is:

$$u^{(1)} = -\frac{(k_1 - k_2)^2 k_1 k_2 \left(\frac{320\alpha_4}{3}\right) g_1 g_2}{\left(1 + g_1 + g_2 + g_1 g_2 \left(\frac{k_1 - k_2}{k_1 + k_2}\right)^2\right)^3} \times$$

$$\left\{1 + \left\{\left(\left(g_1 \frac{k_2}{(k_1 + k_2)}\right)^2 - g_1^2 g_2 \left(\frac{k_2(k_1 - k_2)}{(k_1 + k_2)^2}\right)^2\right) + g_1 \frac{2k_2^2 + 2k_1 k_2 - k_1^2}{(k_1 + k_2)^2} - g_1 g_2 \frac{(k_1^2 - 3k_1 k_2 + k_2^2)}{(k_1 + k_2)^2}\right) + (1 \leftrightarrow 2)\right\}\right\}. \quad (28)$$

In Eq. (28), $g_i$ are defined as in Eqs. (4) and (6). By construction, $u^{(1)}$, is a solution of Eq. (22). Direct substitution confirms this.

The first observation is that $u^{(1)}$ of Eq. (28) vanishes identically in the single-soliton limit, $k_2 = k_1$. (All higher corrections contained in the exact two-soliton solution vanish in this limit, because $A$ in Eq. (7) is proportional to $(k_2 - k_1)^2$). Next, the asymptotic limits of $u^{(1)}$, obtained readily from Eq. (28) ($k_2 > k_1$ is assumed) are:

$$u^{(1)} \to \begin{cases} -\frac{20}{3}\alpha_4 k_1 \partial_x u^{Single}(t, x + \delta_{2,0}; k_2) & ,t \to -\infty \\ -\frac{20}{3}\alpha_4 k_2 \partial_x u^{Single}(t, x + \delta_{1,0}; k_1) & ,t \to +\infty \end{cases}. \quad (29)$$

From Eq. (29), it is obvious that $u^{(1)}$ does not have the structure of the particular solution, $u_{in}^{(1)}$, obtained in the Normal Form expansion (see Eq. (26)). It is not an inelastic collision between a soliton and an anti-soliton. Rather, it is the inelastic "bending" process, in which the incoming state is the faster soliton. It propagates until it hits the localized "scattering potential" term of Eq. (22), which "absorbs" that soliton, and "emits" the slower soliton. This is the characteristic of the full effect of the perturbation shown in Fig. 2.

The asymptotic structure of $u^{(1)}$ in the exact two-soliton solution determines $S$, the linear combination of symmetries that has to be added to the particular Normal Form solution $u_{in}^{(1)}$ in Eq. (23). With two solitons, the first two symmetries are sufficient. They are [10-14,16-23]:

$$S_1 = u_1 \rightarrow \begin{cases} \partial_x \left( u^{Single}(t,x;k_1) + u^{Single}(t,x+\delta_{2,0};k_2) \right) & t \rightarrow -\infty \\ \partial_x \left( u^{Single}(t,x+\delta_{1,0};k_1) + u^{Single}(t,x;k_2) \right) & t \rightarrow +\infty \end{cases}$$

$$S_2 = 6uu_1 + u_3 \rightarrow \begin{cases} \partial_x \left( 4k_1^2 u^{Single}(t,x;k_1) + 4k_2^2 u^{Single}(t,x+\delta_{2,0};k_2) \right) & t \rightarrow -\infty \\ \partial_x \left( 4k_1^2 u^{Single}(t,x+\delta_{1,0};k_1) + 4k_2^2 u^{Single}(t,x;k_2) \right) & t \rightarrow +\infty \end{cases}$$ (30)

For $u^{(1)}$ to have the asymptotic form given by Eq. (29), the linear combination $S$ needs to be

$$S = \frac{5\alpha_4}{6(k_1+k_2)} \left( -4(k_1^2 + k_1 k_2 + k_2^2) S_1 + S_2 \right) .$$ (31)

This completes the construction of $u^{(1)}$ in closed form:

$$u^{(1)} = -\frac{5}{3}\alpha_4 \left( u^2 + q u_1 + u_2 \right) + \frac{5\alpha_4}{6(k_1+k_2)} \left( -4(k_1^2 + k_1 k_2 + k_2^2) S_1 + S_2 \right) .$$ (32)

### 3.3 Higher-order corrections in solution of Eq. (1)

In the Normal Form expansion, the dynamical equations for $u^{(k)}$, $k \geq 2$, have the same structure as Eq. (22). $u^{(1)}$ is replaced by $u^{(k)}$, and the first-order driving term – by an appropriate higher-order driving term. However, unlike in the first-order analysis, there is an enormous level of freedom in higher orders of the expansion. As a result, the driving terms for $k \geq 2$ need not be localized special polynomials. Luckily, this freedom enables one to shape the driving terms in the higher-order dynamical equations so that, like the driving term in Eq. (22), they do vanish identically in the single-soliton case, and are localized around the soliton collision region in the multiple-soliton case. This has been shown for $k = 2, 3$ [15]. With this achieved, it is found that, in the case of Eq. (1), not only $u_{el}^{(1)}$ vanishes identically but so do the elastic components, $u_{el}^{(2)}$ and $u_{el}^{(3)}$. Thus, $u^{(2)}$ and $u^{(3)}$ also represent pure inelastic process. They contain particular solutions (denoted by $u_{in}^{(k)}$), which vanish identically in the single-soliton limit, and possibly, linear combinations of symmetries of the KdV equation. The expressions for $u_{in}^{(2)}$ and $u_{in}^{(3)}$, derived in [15], are reduced to:

$$u_{in}^{(2)} = \frac{1}{2!}\left(-\frac{5}{3}\alpha_4\right)^2 \{14u^3 + 6q^{(3)}u_1 + 18quu_1 + 30u_1^2 + q^2u_2 + 34uu_2 + 4qu_3 + 5u_4\}, \quad (33)$$

and

$$u_{in}^{(3)} = \frac{1}{3!}\left(-\frac{5}{3}\alpha_4\right)^3 \begin{cases} 32u^4 + 30(2q^{(5,1)} - q^{(5,2)})u_1 + 306q^{(3)}uu_1 + 468qu^2u_1 + 45q^2u_1^2 \\ + 2574uu_1^2 + 18qq^{(3)}u_2 + 48q^2uu_2 + 1488u^2u_2 + 516qu_1u_2 \\ + 762u_2^2 + q^3u_3 + 60q^{(3)}u_3 + 222quu_3 + 1167u_1u_3 + 9q^2u_4 \\ + 432uu_4 + 27qu_5 + 39u_6 \end{cases}. \quad (34)$$

In Eqs.(33) and (34):

$$q^{(3)} = \frac{1}{2}\left\{\int_{-\infty}^{x} u^2 dx - \int_{x}^{\infty} u^2 dx\right\}, \quad q^{(5,1)} = \frac{1}{2}\left\{\int_{-\infty}^{x} u^3 dx - \int_{x}^{\infty} u^3 dx\right\}, \quad q^{(5,2)} = \frac{1}{2}\left\{\int_{-\infty}^{x} u_1^2 dx - \int_{x}^{\infty} u_1^2 dx\right\}. \quad (35)$$

The superscripts in $q^{(k)}$ specify their scaling weight [21]. Substitution of the single-soliton solution, Eq. (5) in Eqs. (33) - (35) reveals that, just like $u_{in}^{(1)}$, $u_{in}^{(2)}$ and $u_{in}^{(3)}$ vanish identically in the single-soliton limit and represent pure inelastic processes. For example,

$$u^{(2)}_{in}(t,x) = \underset{|t|\to\infty}{\to} \frac{1}{2!}\left(-\frac{5}{3}\alpha_4\right)^2 \partial_x \sum_{i=1}^{N} \left\{ \begin{array}{l} \left(2Q_i^{(3)} u^{Single}(t,x+\delta_i;k_i)\right) + 9Q_i^{(1)} u^{Single}(t,x+\delta_i;k_i)^2 \\ + \frac{1}{4}Q_i^{(2)} \partial_x u^{Single}(t,x+\delta_i;k_i) + 4Q_i^{(1)} \partial_x^2 u^{Single}(t,x+\delta_i;k_i) \end{array} \right\} . \quad (36)$$

In Eq. (36), the phase shifts obtain their KdV values and

$$Q_i^{(2)} = \sum_{j\neq i}(2k_j)^2 \quad , \quad Q_i^{(3)} = \left\{\sum_{k_j<k_i}(-2k_j)^3 + \sum_{k_j>k_i}(2k_j)^3\right\}\mathrm{sgn}(t) \quad . \quad (37)$$

Thus, $u^{(2)}_{in}$ asymptotes into a sum of well separated single KdV-type soliton solutions of the Normal Form, Eq. (12), each of which is multiplied by the wave numbers of the *other* solitons.

In summary, the Normal-Form expressions for $u^{(2)}$ and $u^{(3)}$ share the properties identified in the exact two-soliton solution of Eq. (1), given by Eqs. (2), (4), (6), (7) and (25). All higher-order corrections contained in the exact two-soliton solution vanish in the single-soliton limit ($k_2 = k_1$), because $A$ of Eq. (7) is proportional to $(k_1 - k_2)^2$. In addition, just like $u^{(1)}$, the higher-order corrections contribute to the inelastic scattering process depicted in Fig. 2. Thus, appropriate linear combinations of the symmetries $S_1$ and $S_2$ must be added to $u^{(k)}_{in}$, $k = 2, 3$. The procedure is identical to the one outlined in the first-order case, hence, is not repeated here.

## 4. Summary

In this paper, the Normal Form asymptotic expansion of the solution of the KdV-Sawada Kotera equation has been compared against, the exact single- and two-soliton solutions of this equation. Both in the Normal Form analysis and in the "template", the higher-order corrections vanish identically in the single-soliton limit, and represent purely inelastic processes. The structure of the exact two-soliton solution of Eq. (1) enables one to compute the linear combination of symmetries that needs to be added to a standard inelastic scattering process in every order of the Normal Form expansion. Whereas it is not known yet whether the KdV-Sawada-Kotera equation has $N$-soliton solutions for any $N$, it is asymptotically integrable, at least through third order.

Figure captions

Fig. 1 – Exact two-soliton solution of Eq. (1) $\alpha_4 = 1$, $\varepsilon = 0.1$, $k_1 = 0.2$, $k_2 = 0.4$

Fig. 2 – Inelastic component above KdV soliton approximation: $\alpha_4 = 1$, $\varepsilon = 0.1$, $k_1 = 0.2$, $k_2 = 0.4$

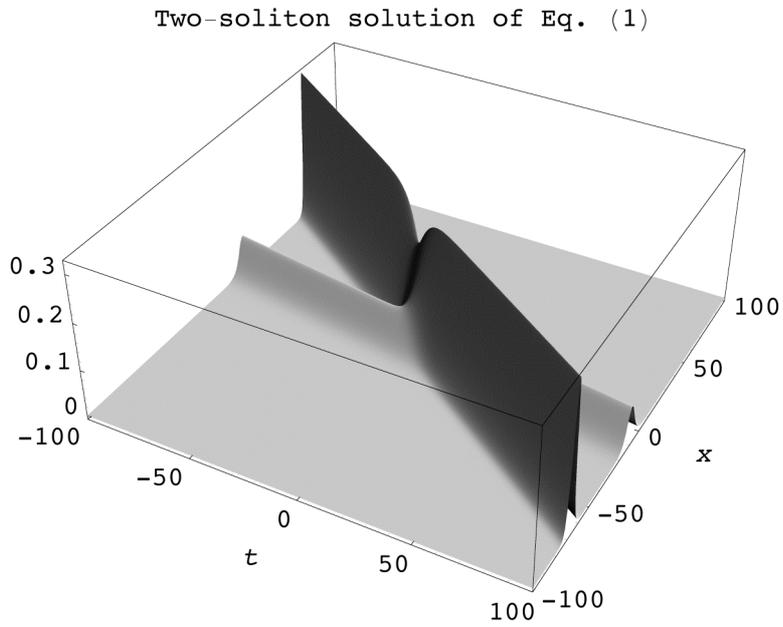

Fig. 1

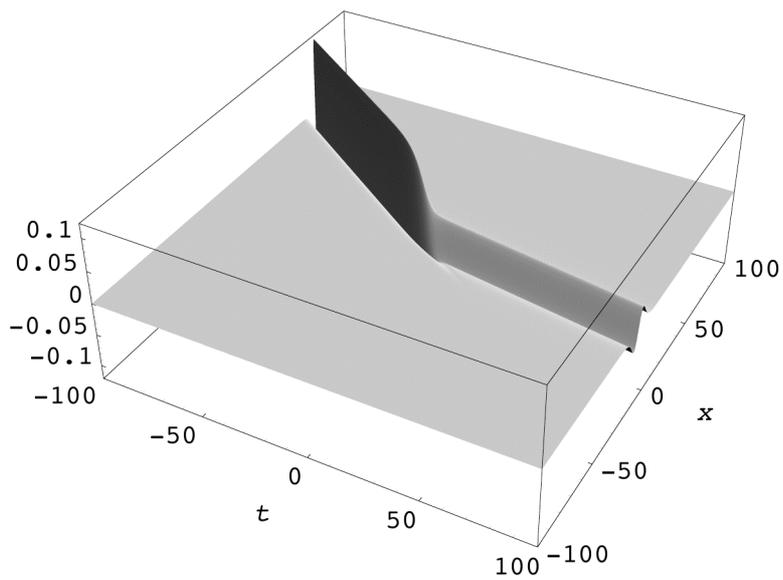

Fig. 2